\documentclass[english,aps,prl, twocolumn]{revtex4-1}
\usepackage[T1]{fontenc}
\usepackage[latin9]{inputenc}
\usepackage{amsbsy}
\usepackage{amssymb}
\usepackage{amsmath}
\usepackage{esint}
\usepackage{graphicx}
\makeatletter
 
 \@ifundefined{textcolor}{}
 {%
   \definecolor{BLACK}{gray}{0}
   \definecolor{WHITE}{gray}{1}
   \definecolor{RED}{rgb}{1,0,0}
   \definecolor{GREEN}{rgb}{0,1,0}
   \definecolor{BLUE}{rgb}{0,0,1}
   \definecolor{CYAN}{cmyk}{1,0,0,0}
   \definecolor{MAGENTA}{cmyk}{0,1,0,0}
   \definecolor{YELLOW}{cmyk}{0,0,1,0}
 }

\makeatother

\usepackage{babel}
\begin{document}

\title{Observing Dirac's classical phase space analog to the quantum state}

\author{Charles Bamber}

\email{charles.bamber@nrc-cnrc.gc.ca}

\affiliation{Measurement Science and Standards, National Research Council, 1200
Montreal Rd, Ottawa, Canada }

\author{Jeff S. Lundeen}

\email{jlundeen@uottawa.ca}

\affiliation{Physics Dept., University of Ottawa, 150 Louis Pasteur, MacDonald
Hall, Ottawa, Canada }

\date{\today}
\begin{abstract}
In 1945, Dirac attempted to develop a ``formal probability\textquotedblright{}
distribution to describe quantum operators in terms of two non-commuting
variables, such as position $x$ and momentum $p$ {[}Rev. Mod. Phys.
17, 195 (1945){]}. The resulting quasi-probability distribution is
a complete representation of the quantum state and can be observed
directly in experiments. We measure Dirac\textquoteright{}s distribution
for the quantum state of the transverse degree of freedom of a photon
by weakly measuring transverse $x$ so as to not randomize the subsequent
$p$ measurement. Further, we show that the distribution has the classical-like
feature that it transforms (e.g., propagates) according to Bayes'
law.
\end{abstract}


\maketitle
While formulating quantum theory early in the century, many physicists
sought a classical interpretation of the object at its center, the
quantum state. The most well-known of these is the Wigner function
$W(x,p)$, an attempt to produce a joint probability distribution
for a particle's instantaneous momentum and position \citep{Wigner1932}.
These `phase-space' distributions necessarily
violate many of the properties that classical probability distributions
must obey. However, they are useful for visualizing concurrent momentum
and position features in quantum states, which might be obscurely
encoded in the phase of the wavefunction or the off-diagonal elements
of the density matrix. Additionally, a non-classical hallmark in the
distribution (e.g. negative probabilities) can be used to identify
intrinsically quantum states \citep{Hudson1974}. It is remarkable
that even though the quantum state has been an overwhelmingly successful
concept and tool for almost a century, our understanding of its nature
is still being refined \citep{Harrigan2010,*Colbeck2011,*Pusey2012,*Colbeck2012,*Lewis2012}.
These distributions have contributed to this refinement \citep{Banaszek1999a,*Brida2002,*Ferrie2008,*Spekkens2008,*Son2009,*Wallman2012}
and have helped demarcate the boundary between classical and quantum
phenomena \citep{Mari2012,*Veitch2012}.

In quantum physics, Heisenberg's uncertainty relation implies that
a precise joint measurement of position $X$ and momentum $P$ is
impossible. Contrast this with classical physics, in which a particle
has a definite and unique position $x$ and momentum $p$ at any moment
in time, thereby defining its `state'. Measuring a classical particle's
state then just entails a joint measurement of $X$ and $P$. Equivalently,
one can measure whether a particle is at a particular point $(x,p)$
in `phase-space' (i.e. $X-P$ space) and then raster over $\mathit{x}$
and $\mathit{p}$. And, if the particle is produced in a random process
such that it is in a random distribution of states, then repeated
measurements at each point will let us find the average result: the
probability for the particle to be at that point, $\mathbb{P}(x,p)$,
a phase-space probability distribution.

Consider what the quantum version of this measurement would be by
beginning with the classical description of this phase-space point,
a two-dimensional Dirac delta function centered at $x$ and $p$,
$\delta^{\left(2\right)}\left(X-x,P-p\right)$. Crucially, there is
no unique nor general method for translating a classical observable
to its quantum equivalent \citep{Groenewold1946}. For example, since
they do not commute one must choose an ordering $O$ of quantum operators
$\mathbf{X}$ and $\mathbf{P}$ with which to replace classical variables
$X$ and $P$: 
\begin{equation}
\mathbf{\boldsymbol{\Delta}_{\mathrm{\mathit{O}}}}(x,p)=\left\{ \delta^{\left(2\right)}\left(\mathbf{X}-x,\mathbf{P}-p\right)\right\} _{O}.
\end{equation}
In this Letter, we experimentally demonstrate the measurement of this
operator for the anti-standard ordering (i.e.$\mathbf{P}$ is always
to the left of $\mathbf{X}$), $\mathbf{\boldsymbol{\Delta}_{\mathrm{\mathit{AS}}}}(x,p)=\delta(\mathbf{P}-p)\delta(\mathbf{X}-x)=\boldsymbol{\pi}_{p}\boldsymbol{\pi}_{x}$,
where $\boldsymbol{\pi}_{m}=\left|m\right\rangle \left\langle m\right|$
is a projector. Numerous other orderings are possible and each corresponds
to a distinct point operator $\boldsymbol{\Delta}_{\mathrm{\mathit{O}}}$,
which may or may not describe a physical measurement (an `observable').
The average result of such a measurement then would be the quantum
version of our classical state measurement procedure outlined above.

As usual, this average result is equal to the expectation value of
the operator, 
\begin{equation}
\left\langle \boldsymbol{\Delta}_{O}(x,p)\right\rangle =\mathrm{Tr}\left[\boldsymbol{\Delta}_{O}(x,p)\boldsymbol{\rho}\right],
\end{equation}
where $\boldsymbol{\rho}$ is the density operator describing the
quantum state of the particle. It may come as a surprise that this
simple, classically motivated measurement procedure will completely
determine the quantum state. Whereas classically it gives the probability
$\mathbb{P}(x,p)$, the quantum version gives a `quasi-probability'
$\mathrm{\mathbb{\widetilde{P}}_{\mathit{\bar{O}}}(x,p)=\left\langle \boldsymbol{\Delta}_{\mathit{O}}\right\rangle }$
\citep{Agarwal1970, Agarwal1970a}, where $\bar{O}$ is the reverse
ordering to $O$. That is, for typical orderings $O$, the average
result is a phase-space quasi-probability distribution in $x$ and
$p$ equivalent to the state of the particle. From this perspective,
the Wigner function corresponds to a direct measurement of an $(x,p)$
point observable that has been symmetrically ordered (i.e. the `Weyl'
ordering $\mathrm{W}$, with $W=\bar{W}$): $\mathrm{\mathbf{\boldsymbol{\Delta}_{\mathit{W}}=R}}_{\pi}(x,p)/\pi$
\citep{Royer1977}, where $\mathbf{R}_{\pi}$ is the parity of a particle
about point $(x,p)$. The normal $N$ ordering ($\mathbf{a^{\dagger}}$
to the left of $\mathbf{a}$, where $\mathbf{a}$ is the usual lowering
operator $\mathbf{a}=\mathbf{X}+i\mathbf{P}$ ) and its reverse, anti-normal
$AN$, correspond to the other two well-known quasi-probability distributions,
the Husimi Q function \citep{Husimi1940} ($\mathbb{\widetilde{P}}{}_{\mathit{N}}$,$\mathbf{\boldsymbol{\Delta}_{\mathrm{\mathit{AN}}}}(\alpha=x+ip)=\left|\alpha\right\rangle \left\langle \alpha\right|/\pi$,
i.e. a projection onto a coherent state $\left|\alpha\right\rangle $
\citep{Agarwal1970} ) and the Glauber-Sudarshan P distribution \citep{Glauber1963,*Sudarshan1963}
($\mathbb{\widetilde{P}}_{\mathit{AN}}$, $\mathbf{\boldsymbol{\Delta}_{\mathrm{\mathit{N}}}}$
is unphysical ). The Wigner and Q functions have been directly measured
in various physical systems \citep{Vogel1989,*Banaszek1996,*Leonhardt1997,*Lutterbach1997,*Bertet2002,*Kanem2005,*Hofheinz2009,*Laiho2010},
including the transverse state of a photon \citep{Mukamel2003,*Smith2005}.

\begin{figure}[h]
\includegraphics[width=3.6in,bb=1 1 3840 2160]{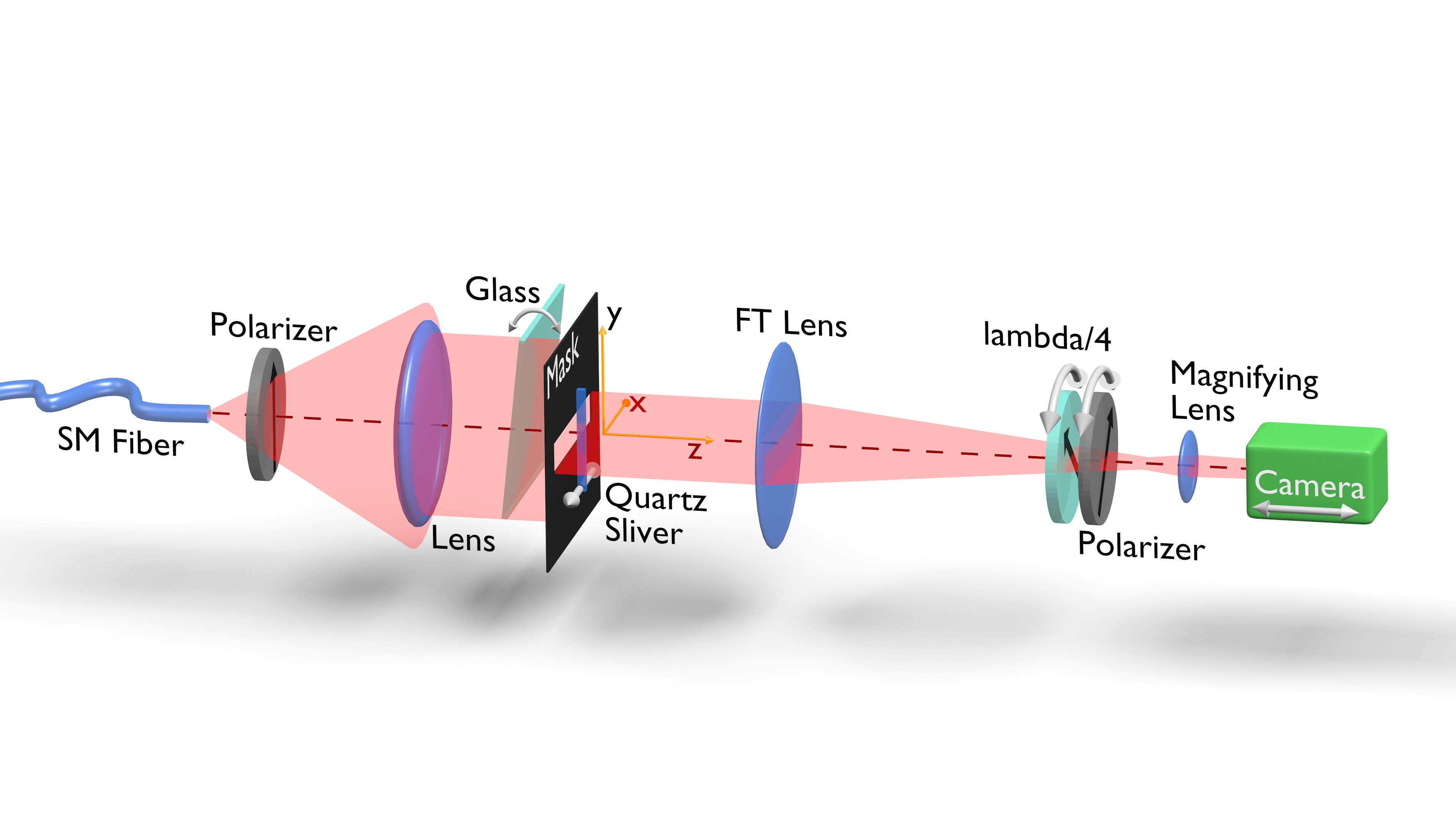}
\caption{\label{Figure_1} Experimental setup to directly measure the
Dirac quasi-probability distribution $\mathrm{\mathbb{\widetilde{P}}}(x,p)$
of the transverse state of a photon. Photons emerge from a single
mode fiber and are collimated to produce an identical ensemble in
an initial pure state. This is optionally transformed to a mixed state
by noise introduced by an oscillating glass plate. A weak measurement
of transverse position $\boldsymbol{\pi}_{x}=\left|x\right\rangle \left\langle x\right|$
is implemented by a small polarization rotation $\varphi=12.92^{\circ}$
created by a quartz sliver at $x$. This weak measurement is read
out jointly with the transverse momentum $p$ by using a camera, polarizer
and quarter waveplate in the Fourier transform plane to measure $\left\langle \boldsymbol{\pi}_{p}\boldsymbol{\sigma}^{-}\right\rangle =\sin\varphi\mathrm{\mathbb{\widetilde{P}}}(x,p)$.
We repeat this measurement after displacing the camera from the Fourier
transform plane by $\Delta z$ in order to investigate how the distribution
propagates in space.}
\end{figure} 

In 1945, Dirac wrote ``On the analogy between classical and quantum
mechanics,'' \citep{Dirac1945} in which he introduced what we call
the `Dirac distribution' as a classical-like representation of a quantum
operator, such as the density operator $\boldsymbol{\rho}$. Over
the past 90 years this distribution has been repeatedly rediscovered
\citep{McCoy1932,Kirkwood1933,*Terletsky1937,*Rivier1951,*Barut1957,*Margenau1961,*Levin1964,*Rihaczek1968,*Ackroyd1970}.
It has since been realized that the Dirac distribution is $\mathrm{\mathbb{\widetilde{P}}}_{\mathit{S}}(x,p)$,
the standard ordered quasi-probability distribution and its complex
conjugate is $\mathbb{\widetilde{P}}_{A\mathit{S}}(x,p)$ \citep{Agarwal1970,Praxmeyer2003,Chaturvedi2006}.
Despite being the product of one of the founders of quantum theory,
there has been little experimental investigation of the Dirac distribution
over the past 90 years. The impediment has been that although its
direct measurement appears simple, i.e. a measurement of position
then momentum, one quickly notices that $\mathbf{\boldsymbol{\Delta}_{\mathrm{\mathit{AS}}}}=\boldsymbol{\pi}_{p}\boldsymbol{\pi}_{x}$
is non-Hermitian and thus should not be an observable. Physically,
this is enforced by the fact that a measurement of $\boldsymbol{\pi}_{x}=\delta(\mathbf{X}-x)$
will disturb and hence invalidate the subsequent measurement of $\boldsymbol{\pi}_{p}=\delta(\mathbf{P}-p)$,
making a joint measurement impossible. 

Two ways in which measurement-induced disturbance can be minimized
in quantum physics are: 1. by lowering the precision, $\delta(x)\rightarrow$$\Delta x$,
and 2. by decreasing the certainty of the measurement,$\mathrm{Prob}(x_{system}|x_{measured})\ll1$.
An example of the second approach is \textit{weak} measurement: by
reducing the coupling between the system and the measurement apparatus,
the result from any one trial becomes uncertain \citep{Pryde2005,*Resch2004,*Mir2007,*Williams2008,*Hosten2008,*Lundeen2009,*Yokota2009,*Dixon2009,*Feizpour2011,*Kocsis2011}.
Reducing the coupling similarly reduces the disturbance (i.e. back-action).
Moreover, by averaging over many trials an \textit{average} result
can still be found to arbitrarily low uncertainty. Remarkably, the
Dirac distribution can be measured simply by replacing the first measurement
by a \textit{weak} measurement \citep{Aharonov1988} of $\boldsymbol{\pi}_{x}$,
as we showed in Ref. \citep{Lundeen2012} (see also the related work
\citep{Arthurs1965,*Johansen2008,*Kalev2012,*DiLorenzo2013,*Wu2013}).We
termed the average joint result of this weak-strong position-momentum
measurement, the `weak average'. In the zero-coupling limit, it
is \citep{Lundeen2012}, 
\begin{equation}
\left\langle \boldsymbol{\pi}_{p}^{\mathrm{s}}\boldsymbol{\pi}_{x}^{\mathrm{w}}\right\rangle _{\rho}=\mathrm{Tr}\left[\boldsymbol{\pi}_{p}\boldsymbol{\pi}_{x}\boldsymbol{\rho}\right]=\left\langle \mathbf{\boldsymbol{\Delta}_{\mathrm{\mathit{AS}}}}(x,p)\right\rangle =\mathrm{\mathbb{\widetilde{P}}}_{\mathit{S}}(x,p),
\end{equation}
where $\mathrm{\mathbb{\widetilde{P}}}_{\mathit{S}}$ is the Dirac
distribution (see Supp. Mat.). The superscripts s and w denote strong
(i.e. normal) and weak measurements, respectively. (From here on,
we omit the $S$ subscript as we will deal exclusively with the Dirac
distribution.) As an expectation value of a non-Hermitian operator,
the weak average is complex in general \citep{Jozsa2007}. The real
component manifests itself in the usual pointing variable of the measurement
apparatus while the imaginary component manifests itself in the conjugate
variable, as we will describe later using our specific experiment
as an example. It follows that, unlike the Wigner function and other
aforementioned quasi-probabilities, the Dirac distribution is complex. 

The Dirac distribution $\mathrm{\mathbb{\widetilde{P}}}(x,p)$ possesses
the key feature that it can be manipulated according Bayes' Theorem
despite the fact that it is not a true probability \citep{Steinberg1995a,Hofmann2012}.
For example, we could calculate the conditional quasi-probability
of $x$ given $p$: $\mathrm{\mathbb{\widetilde{P}}}(x|p)=\mathrm{\mathbb{\widetilde{P}}}(x,p)/\mathrm{\mathbb{\widetilde{P}}}(p)$,
where $\mathrm{\mathbb{\widetilde{P}}}(p)\equiv\left\langle \boldsymbol{\pi}_{p}^{\mathrm{w}}\right\rangle _{\rho}=\mathbb{P}(p)$
is defined in analogy to the weak average. One could also directly
\textit{measure} $\mathrm{\mathbb{\widetilde{P}}}(x|p)$ by following
our naive procedure above but only keeping the results for $X$ in
those cases where $P=p$. For $p=0$, this is a succinct description
of our previously introduced procedure to directly measure the quantum
wavefunction $\Psi(x)$ \citep{Lundeen2011}. In this light, one would
write $\Psi(x)\propto\mathrm{\mathbb{\widetilde{P}}}(x|p=0)$, which
provides wavefunction with a pithy description: It is the quasi-probability
of $x$ given that $p$ was found to be zero.

\begin{figure}
\includegraphics[width = 3.375in]{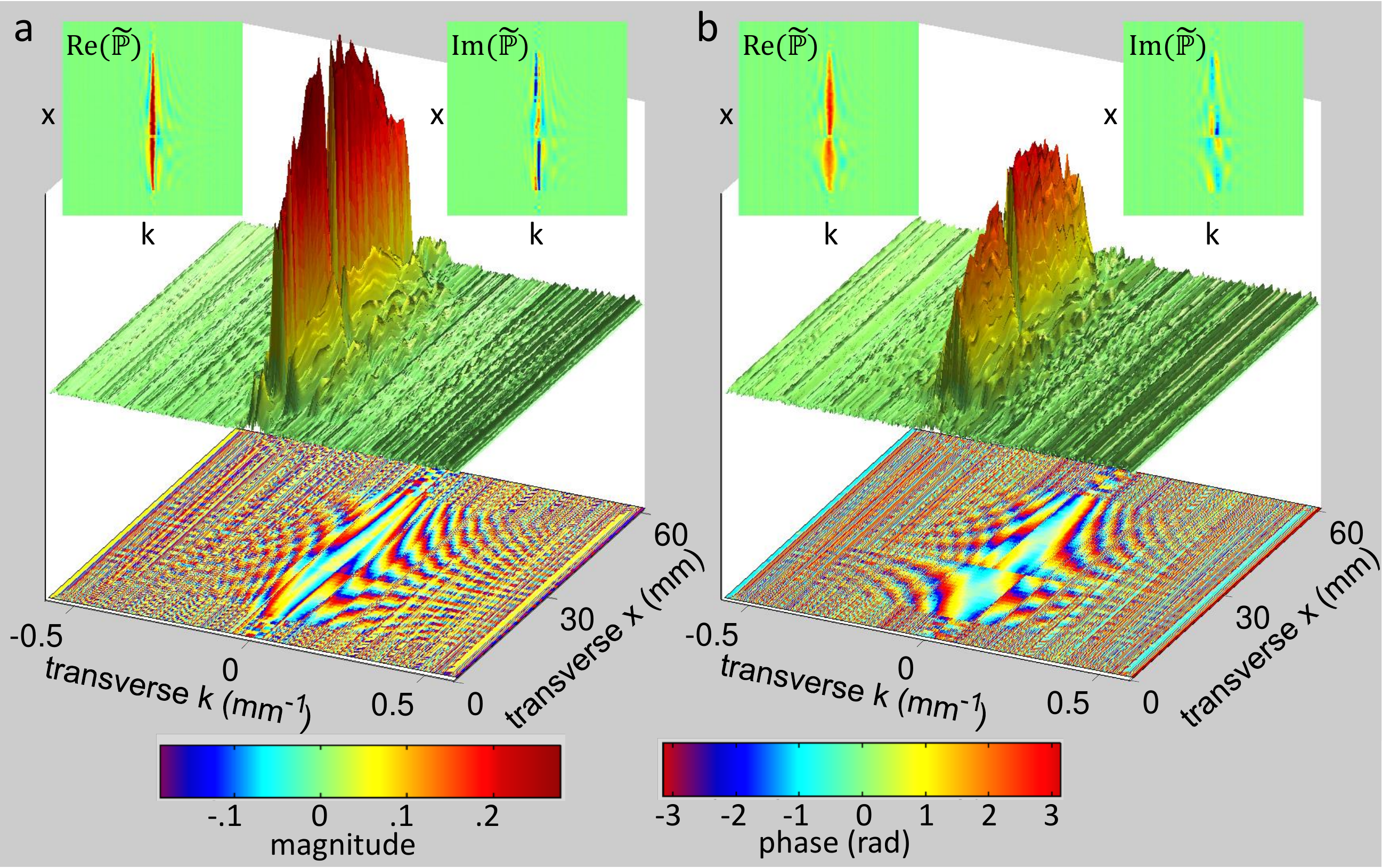}
\caption{\label{Figure_2} Directly measured Dirac distributions for
a (a) pure and a (b) mixed state. In both (a) and (b), the left and
right insets are the directly measured real and imaginary parts of
the Dirac distribution $\mathrm{\mathbb{\widetilde{P}}}(x,p)$. From
these we calculate the magnitude (upper plot) and phase (lower plot)
of the distribution. In the axes, transverse $x$ corresponds to the
position of the quartz sliver and, thus, the weak measurement of $\left|x\right\rangle \left\langle x\right|$;
transverse $k=p/\hbar$ is the transverse wavevector, which is proportional
to the transverse position in the Fourier transform plane at the camera
sensor (see text).}
\end{figure} 

Unlike our wavefunction measurement procedure \citep{Lundeen2011},
the measurement of the Dirac distribution can also determine the state
of a system that is `mixed'. In this case, the state is fully characterized
by a density operator $\boldsymbol{\rho}$ rather than a wavefunction
$\Psi(x)$. The density operator is a generalization of the wavefunction
(i.e. a `pure' state) that can incorporate classical noise and the
effects of entanglement with other systems. It is widely used in statistical
mechanics, quantum chemistry, quantum information and atomic physics.

As an example, we measure the Dirac distribution of the quantum state
corresponding to the transverse position of a photon for mixed and
pure states. We clarify what we mean by the photon's quantum state
in the Supplementary Material. Recent work has measured the Dirac
distribution in a discrete system \citep{Salvail2013} but only for
the pure case, and only in a simple two-level system. Shown in Fig.
1, our experimental setup builds on the one in Ref. \citep{Lundeen2011}.
Our photons are produced by an attenuated laser (wavelength=780 nm)
and coupled into a single-mode fiber. Although we do not use single-photon
states, one can say that every photon that exits the fiber output
will have the same transverse state; they form an identical ensemble
of particles. The photons are linearly polarized and collimated by
a convex lens (achromat, focal length $f=30$ cm, diameter $d=5$
cm) and sent through an aperture (x$\times$y dimensions=44 mm$\times$2
mm). Unlike in Ref. \citep{Lundeen2011}, just before the lens, we
introduce phase-noise by rotating a 4 mm thick glass plate by 4 degrees
about a horizontal rotation axis at 11 Hz, thereby generating many
waves of phase delay. The plate extends only over one half of the
transverse state. With the glass stationary, the transverse state's
phase is discontinuous at the position of the plate edge. With it
oscillating, the two halves of the state are completely incoherent
over the time-scale of our measurements and, thus, the state is mixed.

We divide the weak measurement of $\boldsymbol{\pi}_{x}=\left|x\right\rangle \left\langle x\right|$
into two stages: coupling and readout. The coupling stage occurs just
after the collimating lens. This is the plane in space at which we
measure the Dirac distribution of the photon ensemble. Here, a quartz
sliver (width $\Delta x=1$ mm) rotates the photon polarization (initially
$0^{\circ}$) to $\varphi$ degrees at position $x$. For $\varphi=90^{\circ}$
this would be a strong measurement, whereas by setting $\varphi\ll90^{\circ}$
our measurement becomes weak. The sliver is also slightly tilted about
the x-axis in order to null any phase shift it induces. As described
in Ref. \citep{Lundeen2011}, we can readout by measuring $\left\langle \boldsymbol{\pi}_{x}^{\mathrm{w}}\right\rangle =(\left\langle s\right|\boldsymbol{\sigma}_{x}\left|s\right\rangle -i\left\langle s\right||\boldsymbol{\sigma}_{y}\left|s\right\rangle )/\sin\varphi=\left\langle s\right|\boldsymbol{\sigma^{-}}\left|s\right\rangle /\sin\varphi$,
where $\left|s\right\rangle $ is the polarization state of the photon,
$\boldsymbol{\sigma}_{x}$ and $\boldsymbol{\sigma}_{y}$ are the
Pauli operators, and $\boldsymbol{\sigma}^{-}=\boldsymbol{\sigma}_{x}+i\boldsymbol{\sigma^{-}}$
is the lowering operator \citep{Lundeen2005}. The real part and imaginary
parts of the weak value are proportional to the shift from zero of
the average value of $\boldsymbol{\sigma}_{x}$ and $\boldsymbol{\sigma}_{y}$,
respectively. Thus, as expected the two parts separately appear in
conjugate variables of our measurement apparatus, that is, in the
linear and circular polarizations.

\begin{figure}[h]
\includegraphics[width = 3.5in]{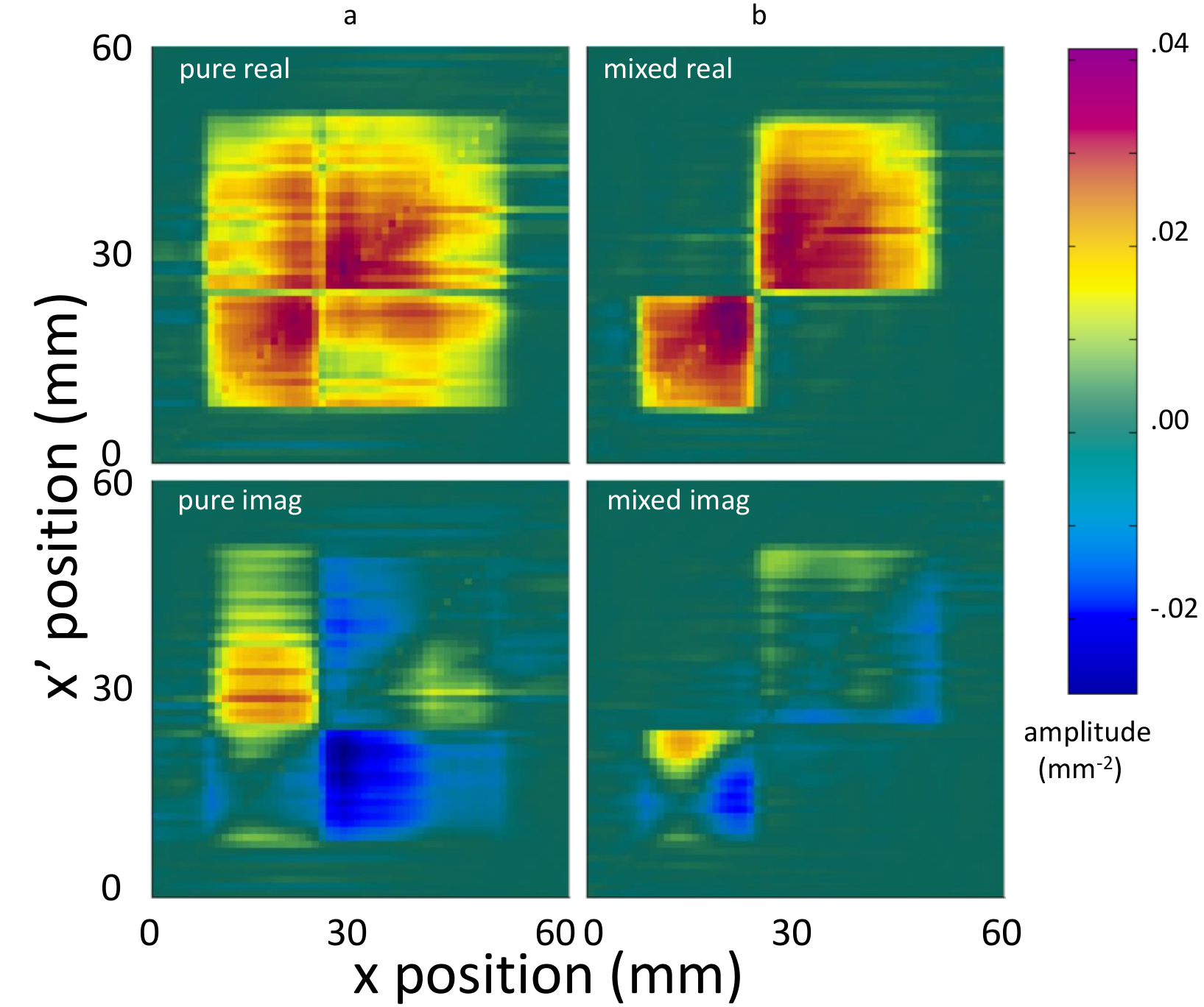}
\caption{\label{Figure_3} Density matrices $\rho$ for a (left plots)
pure and a (right plots) mixed state, calculated from data presented
in Fig. 2. The upper plots and lower plots correspond to the real
and imaginary parts of $\rho$, respectively.We have divided the diagonals
by $\cos(\varphi)$ to correct for effect of the measurement back-action
(see the Supp. Mat. for details)}
\end{figure} 

In order to perform a \textit{joint} measurement of $\boldsymbol{\pi}_{p}\boldsymbol{\pi}_{x}$
we must make polarization measurements at each momentum. To do so,
a lens (achromat, $f_{FT}=$1 m, $d=$5 cm) is placed one focal length
after the sliver. The Fourier transform (FT) of the quantum state
forms in the plane one focal length past the lens. Consequently, the
transverse position $x_{FT}$ in this FT plane is proportional to
the transverse momentum $p$ of the photon at the sliver. We magnify
($M=4.935$) by another lens ($f=35$ mm, $d=2.5$ cm) so that the
final scaling is $p=x_{FT}h/(f_{FT}M\lambda)$, where $h$ is Planck's
constant. To readout $\left\langle \boldsymbol{\pi}_{x}^{\mathrm{w}}\right\rangle $
we project onto a circular or linear polarization by inserting a quarter-waveplate
($\mathrm{lambda}/4$) and/or polarizer (Pol, Nanoparticle Linear
Film Polarizer, Thorlabs LPVIS50) just before the magnifying lens.
Then, position and momentum are jointly measured by measuring $\boldsymbol{\pi}_{p}\boldsymbol{\sigma}^{-}$.
We do so by recording the average number of photons $N_{p,j}$ arriving
at each transverse position $x_{FT}$ on a camera sensor (Basler acA1300-30um,
pixel size: 3.75$\mathrm{\mu m}$$\times$3.75$\mathrm{\mu m}$, x$\times$y
array size: 1296$\times$966, 12 bit) for two pairs of polarization
measurements: $j=45^{\circ}$ and $-45^{\circ}$ polarizations, and
right ($\circlearrowleft$) and left-hand ($\circlearrowright$) circular
polarizations. For $\varphi\ll90^{\circ},$ the differences in each
pair are proportional to the real and imaginary parts of the Dirac
distribution,

\begin{equation}
\left\langle \boldsymbol{\pi}_{p}\boldsymbol{\sigma}^{-}\right\rangle =\frac{N_{p,45^{\circ}}-N_{p,-45^{\circ}}}{N_{p,45^{\circ}}+N_{p,-45^{\circ}}}-i\frac{N_{p,\circlearrowleft}-N_{p,\circlearrowright}}{N_{p,\circlearrowleft}+N_{p,\circlearrowright}}=\mathrm{\mathbb{\widetilde{P}}}(x,p)\sin\varphi,\label{eq:Exp_Dirac}
\end{equation}
where the expectation value is now an average over the polarization
and transverse momentum state of the photon ensemble. Each polarization
measurement is a 1.8 s camera exposure in which we average along the
$y$ dimension to arrive at a vector $N_{p,j}$. We take the mean
of the weak average over ten scans of $x$.

We directly measure the Dirac distribution $\mathrm{\mathbb{\widetilde{P}}}(x,p)$
for the transverse quantum state by measuring these polarization differences
for every $p$ as a function of the sliver position $x$, which we
move in steps of 1 mm across aperture of the collimating lens. The
insets of Fig. 2 plot this pair of polarization differences as the
real and imaginary parts of $\mathrm{\mathbb{\widetilde{P}}}(x,p)$
according to Eq. \ref{eq:Exp_Dirac}. Fig. 2 (a) and (b) display the
measured Dirac distribution for the case where the glass plate is
stationary (pure state) and oscillating (mixed state), respectively.
As can be seen in the lower plots, there is a state-independent phase
of $\exp(-ixk),$ inherent to the Dirac distribution, imposed on the
underlying 2-d form of the state in (a) and (b). This overlay phase
structure allows one to immediately see phase jumps, such as the one
in the pure state (a) at $x=$25 mm (the glass edge) due to increased
optical path length through the glass. In contrast, in the mixed state
(b), the phase fringes on either side of $x=$25 mm are unrelated,
a signature of the lack of phase coherence across this point. Looking
now at the magnitude, shown in the upper plots, both the mixed and
pure states exhibit a depression at $x=25$ mm, likely due to photons
being scattered out of the apparatus by the glass edge. In the momentum
direction, the width of the mixed state is broader than the pure state,
as is expected due to the reduced spatial coherence of the former.
This is accompanied by a decreased magnitude since the distribution
is normalized to one. These distinctive features suggest that the
Dirac distribution provides a useful way to visualize key characteristics
of pure and mixed quantum states.

The Dirac distribution is related in a simple way to the position
density matrix of the state, $\rho(x,x')=\left\langle x\right|\boldsymbol{\rho}\left|x'\right\rangle =\mathrm{FT}\left[\mathrm{\mathbb{\widetilde{P}}}(x,p)\exp(ipx/h)\right](x')$,
where the Fourier transform FT is performed with respect to momentum
$p$ \citep{Chaturvedi2006}. In Fig. 3, we plot the position density
matrix for each of the Dirac distributions from Fig. 2. For the pure
state Fig. 3 (a), the hard edges of the aperture form a square outline
in $\boldsymbol{\rho}$ and the phase jump now appears at both $x=$25
mm and $x'=$25 mm. Strikingly, in the mixed state in Fig. 3 (b) the
off-diagonal regions for $x$$\leq25$ mm, $x'$>25 mm, and the reverse
are close to zero. This is indicative of the lack of coherence between
the part of the state that passes through the oscillating glass and
the part that does not. This shows that we can successfully measure
the Dirac distribution for a transverse quantum state and that it
correctly determines the state of a mixed system. Since we are not
in the limit of a zero interaction-strength measurement ($\varphi=0$),
there will still be some backaction due to the measurement. While
in the Dirac distribution this leads to minor offsets $\mathrm{\mathbb{\widetilde{P}}}(x,p)\rightarrow\mathrm{\mathbb{\widetilde{P}}}(x,p)-\mathrm{Prob}(x)(1-\cos(\varphi))$
everywhere, in the density matrix it leads solely to a suppression,
by $\cos(\varphi)$, of the diagonals, which we correct for in Fig
3 (see Supp. Mat.).

\begin{figure*}[!ht]
\includegraphics[width =7in]{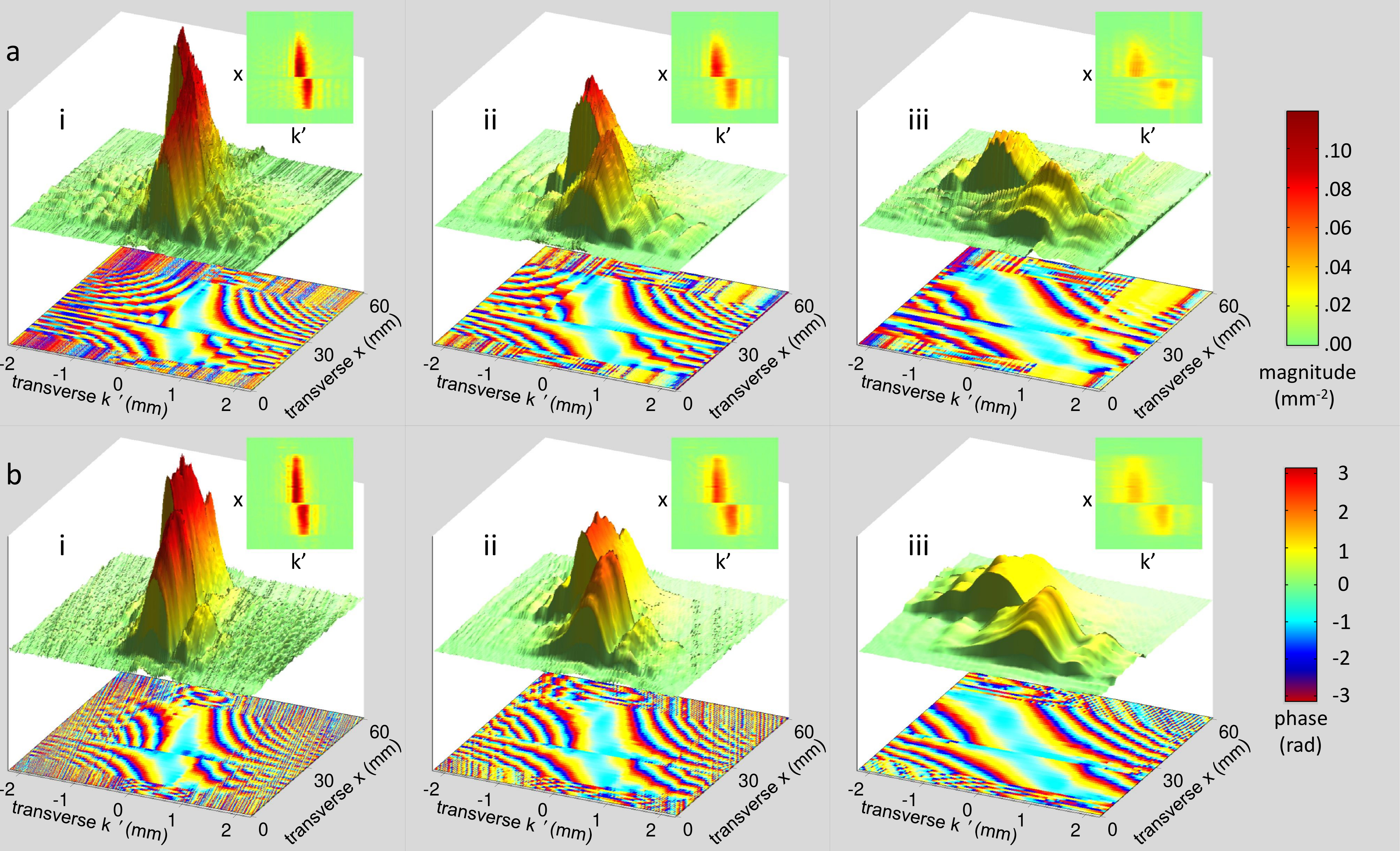}
\caption{\label{Figure_4} Evolution of the quantum state by Bayes'
theorem. (a) A series of directly measured Dirac distributions propagated
from the mixed state in Fig. 2 (b) to a plane translated by $\Delta z=$
(i) 8.4 cm, (ii) 16 cm, and (iii) 32.5 cm past the Fourier transform
plane (by moving the camera). (b) The theoretical prediction for the
propagated distributions found using Bayes' theorem, as in Eq. \ref{eq:Bayes}.
In all (a-b) (i-iii), the bottom plot is Dirac distribution's phase
and both the inset and top plot are its magnitude. As in Fig. 2, the
transverse $x$ axis is the position of the weak measurement, whereas
transverse $k'$ now corresponds to the transverse position on the
camera (which is no longer proportional to momentum).}
\end{figure*} 

The compatibility of Bayes' theorem with the Dirac quasi-probability
distribution goes beyond the simple example we gave earlier. The simplicity
of Dirac distribution allows us to generalize its theoretical definition
to multiple variables, e.g., $\mathrm{\mathbb{\widetilde{P}}}(x,q',k',p)=\left\langle \delta(\mathbf{P}-p)\delta(\mathbf{K'}-k')\delta(\mathbf{Q'}-q')\delta(\mathbf{X}-x)\right\rangle $=$\mathrm{Tr}\left[\boldsymbol{\pi}_{p}\boldsymbol{\pi}_{k'}\boldsymbol{\pi}_{q'}\boldsymbol{\pi}_{x}\boldsymbol{\rho}\right]$
, where $Q'$ and $K'$ are another two continuous variables (e.g.
the photon position and momentum at a later time after undergoing
some evolution). Hoffman showed that with the above theoretical definition
of $\mathrm{\mathbb{\widetilde{P}}}(x,q',k',p)$ one can propagate
the Dirac distribution $\mathrm{\mathbb{\widetilde{P}}}(x,k)$ in
time by applying Bayes' theorem \citep{Hofmann2012}: 
\begin{eqnarray}
\mathrm{\mathbb{\widetilde{P}}}(q',k') & = & \sum_{x,p}\mathrm{\mathbb{\widetilde{P}}}(x,q',k',p)\\
 & = & \sum_{x,p}\mathrm{\mathbb{\widetilde{P}}}(q',k'|x,p)\cdot\mathrm{\mathbb{\widetilde{P}}}(x,p),\label{eq:Bayes}
\end{eqnarray}
where $\mathrm{\mathrm{\mathbb{\widetilde{P}}}}(q',k'|x,p)=\mathrm{\mathbb{\widetilde{P}}}(x,q',k',p)/\mathrm{\mathbb{\widetilde{P}}}(x,p)=\left\langle p|k'\right\rangle \left\langle k'|q'\right\rangle \left\langle q'|x\right\rangle /\left\langle p|x\right\rangle $
is independent of the quantum state. This four dimensional generally
complex conditional quasi-probability is a propagator of an arbitrary
point in $x,p$ phase-space to any point in $q',k'$ phase-space.
In the context of quantum information, it functions like a superoperator
(i.e. it transforms between density operators) and can encompass both
unitary and non-unitary evolution of any quantum process (see \citep{Hofmann2013,*Morita2013}
for further development of this concept). Note that this is very different
from using Bayes' theorem to update a prior quantum state based on
the incomplete information about the system provided by classical
positive-valued statistics, e.g. POVMs (as was studied in \citep{Barnett2000,*Caves2002}).
In particular, here, as with classical probability distributions,
one can directly apply Bayes' theorem to update the Dirac quasi-probability
distribution and evolve it as a function of space or time.

We now describe a demonstration of this Bayesian propagation. For
the sake of experimental simplicity, we only update one of the variables
in the Dirac distribution $\mathrm{\mathrm{\mathbb{\widetilde{P}}}(x,p)}$
to arrive at $\mathrm{\mathrm{\mathbb{\widetilde{P}}}}(x,k')=\left\langle \boldsymbol{\pi}_{k'}^{\mathrm{s}}\boldsymbol{\pi}_{x}^{\mathrm{w}}\right\rangle _{\rho}$.
By moving the camera back from the re-imaged Fourier transform plane
so that the photons travel a further distance $\Delta z$ before being
detected, we change our strong measurement of $P$ to one of $K'$,
which is a combination of $X$ and $P$ \citep{Lohmann1993} (See
Supp Mat.). Then we make a joint weak-strong measurement of $X$ and
$K'$ in exactly the same manner as we did for the Dirac distribution
to experimentally measure $\mathrm{\mathrm{\mathbb{\widetilde{P}}}}(x,k')$.
We repeat the experiment {[}Fig 4. (a){]} and theoretical {[}Fig 4.
(b){]} Bayes' propagation (see Supp. Mat.) for three values of $\Delta z,$
and plot the results in Fig. 4 (i-iii). As $\Delta z$ is increased,
the distributions in Fig. 4 exhibit a broadening of $\mathrm{\mathrm{\mathbb{\widetilde{P}}}}(x,k')$
in the $K'$ direction. This is due to the broad width of the photon
state in $X$ and the fact that the $X$ portion of the hybrid $X,P$
variable $K'$ increases as $\Delta z$ is increased. Also apparent
is a growing $k'$ displacement of the $x>25$mm portion of the state
as $\Delta z$ is increased. This is consistent with a wedge in our
oscillating glass plate of 0.4 arcseconds. Evidently, each of the
three pairs of experimental and theoretical distributions agree well
in a qualitative manner, which confirms the applicability of classical-like
Bayesian propagation.

In conclusion, by experimentally measuring the Dirac quasi-probability
distribution we have completely determined a mixed quantum state.
We have also demonstrated that the Dirac distribution measured at
different spatial planes are related by Bayes' law, which therefore
acts as a propagator of the quasi-probability. Quasi-probability distributions
like the Q, P, or Wigner function reflect the arbitrary choice of
the operator ordering (normal, anti-normal, or symmetric) that embodies
the inherent incompatibility of quantum and classical physics. Missing
from this list has been the standard ordering, the Dirac quasi-probability
distribution, which has three outstanding features: 1. Its measurement
is simple and similar to the classical equivalent. 2. It is compatible
with Bayes' theorem, with which we can propagate it to other points
in space or time. 3. In the limit of a pure quantum state, it reduces
to quantum wavefunction itself.

\appendix

\begin{acknowledgments}
We thank Aephraim Steinberg and Holger F. Hofmann for useful discussions.
\end{acknowledgments}

\section{Supplementary Material for `Observing Dirac's classical phase space
analog to the quantum state'}

\subsection{Weak-strong measurement of joint operators}

In this section, we show that the joint weak-strong measurement of
position-momentum projectors is equal to the average given by the
standard Born rule: $\left\langle \boldsymbol{\pi}_{p}^{\mathrm{s}}\boldsymbol{\pi}_{x}^{\mathrm{w}}\right\rangle _{\rho}=\mathrm{Tr}\left[\boldsymbol{\pi}_{p}\boldsymbol{\pi}_{x}\boldsymbol{\rho}\right]$.
We call this the `weak average'.This result was proven using an
explicit Hamiltonian that models the process of this joint measurement
in our previous Letter \cite{Lundeen2012}. That result was shown
to be true for general observables and for a final measurement that
can be either weak or strong. Here, we give a less rigorous but more
intuitive derivation of this result by restricting ourselves to observables
that are projectors and to a strong final measurement. 

We begin with a review of the `weak value' in this context. One
starts with an ensemble of quantum systems all in an identical initial
state $\left\vert \Psi\right\rangle $. For each member of the ensemble,
one weakly measures a projector $\boldsymbol{\pi}_{a}$ and then strongly
measures another observable $\mathbf{C}$. Consider the sub-ensemble
where that second measurement gave outcome $c$. In the limit of zero-interaction
strength, the average result of measurement $\mathrm{\mathbf{A}}$
in this sub-ensemble is called the weak value \cite{Aharonov1988}:
\[
\left\langle \mathrm{\mathbf{\boldsymbol{\pi}_{a}}}\right\rangle _{\Psi c}=\frac{\left\langle c\right\vert \mathrm{\mathrm{\boldsymbol{\pi}_{a}}}\left\vert \Psi\right\rangle }{\left\langle c|\Psi\right\rangle }.
\]
 This is an average result conditioned (i.e. post-selected) on outcome
$c$. If we instead want $\left\langle \boldsymbol{\pi}_{c}^{\mathrm{s}}\boldsymbol{\pi}_{a}^{\mathrm{w}}\right\rangle _{\Psi}$,
the average result of $\mathrm{\mathbf{\boldsymbol{\pi}_{a}}}$ \textit{and}
$c$, then, as usual, one multiplies the conditional average of $\mathrm{\mathbf{\boldsymbol{\pi}_{a}}}$
by $\mathrm{Prob}(c)=\left\langle c|\Psi\right\rangle \left\langle \Psi|c\right\rangle $:
\begin{eqnarray}
\left\langle \boldsymbol{\pi}_{c}^{\mathrm{s}}\boldsymbol{\pi}_{a}^{\mathrm{w}}\right\rangle _{\Psi} & = & \frac{\left\langle c\right\vert \mathrm{\mathrm{\boldsymbol{\pi}_{a}}}\left\vert \Psi\right\rangle }{\left\langle c|\Psi\right\rangle }\left\langle c|\Psi\right\rangle \left\langle \Psi|c\right\rangle \label{eq:}\\
 & = & \left\langle c\right\vert \mathrm{\mathrm{\boldsymbol{\pi}_{a}}}\left\vert \Psi\right\rangle \left\langle \Psi|c\right\rangle \label{eq:Bayes}
\end{eqnarray}
Now, we want to generalize this result to mixed states, $\boldsymbol{\rho}=\sum_{j}$$\lambda_{j}$$\left|\Psi_{j}\right\rangle $$\left\langle \Psi_{j}\right|$.
Each state in this sum contributes a joint weak-strong average $\left\langle \boldsymbol{\pi}_{c}^{\mathrm{s}}\boldsymbol{\pi}_{a}^{\mathrm{w}}\right\rangle _{\Psi_{j}}$:
\begin{eqnarray*}
\left\langle \boldsymbol{\pi}_{c}^{\mathrm{s}}\boldsymbol{\pi}_{a}^{\mathrm{w}}\right\rangle _{\rho} & = & \sum_{j}\lambda_{j}\left\langle \boldsymbol{\pi}_{c}^{\mathrm{s}}\boldsymbol{\pi}_{a}^{\mathrm{w}}\right\rangle _{\Psi_{j}}\\
 & = & \sum_{j}\lambda_{j}\left\langle c\right\vert \mathrm{\mathrm{\boldsymbol{\pi}_{a}}}\left\vert \Psi_{j}\right\rangle \left\langle \Psi_{j}|c\right\rangle \\
 & = & \left\langle c\right\vert \mathrm{\mathrm{\boldsymbol{\pi}_{a}}}\sum_{j}\lambda_{j}\left\vert \Psi_{j}\right\rangle \left\langle \Psi_{j}|c\right\rangle \\
 & = & \left\langle c\right\vert \mathrm{\mathrm{\boldsymbol{\pi}_{a}}}\boldsymbol{\rho}\left|c\right\rangle \\
 & = & \mathrm{Tr}\left[\boldsymbol{\pi}_{c}\boldsymbol{\pi}_{a}\boldsymbol{\rho}\right].
\end{eqnarray*}
If $a=x$ and $c=p$, this proves our result.

\subsection{The Photon Wavefunction}

Over the last twenty years, the concept of a photon's wavefunction
has been clarified. The photon's full quantum state describes its
freqency-time, position-momentum, and polarization degrees of freedom.
The quantum state formalism has been successfully used in hundreds,
if not thousands, of papers to theoretical describe the physics of
single and entangled photons in experiments, so one might ask why
there is any controversy. Confusion about this concept largely arose
because the concept of the wavefunction was introduced in terms of
the non-relativistic Schr�dinger Equation, yet 1. the Schr�dinger
Equation contains a mass parameter, whereas the photon is massless,
2. The position operator in Schr�dinger Equation is not a true observable
for photons; they are naturally relativistic and cannot be localized
to an exact position \cite{Newton1949}. Compounding this confusion
is the fact that Dirac derived the second quantization (i.e. Quantum
Electrodynamics) for the electromagnetic field \textcolor{black}{before
he considered its first quantization}.

In short, the answer to this confusion is that Maxwell's equations
play the same role to the photon as the Schr�dinger Equation does
to, say, an electron \cite{Sipe1995,Bialynicki-Birula1996,Smith2007}.
\textcolor{black}{The solution to these equations }is the first quantization
of the electromagnetic field, the wavefunction of the photon. In fact,
Maxwell's Equations can be combined into a single equation that has
the same analytic form as the Schr�dinger Equation. This resolves
issue number 1; the Schr�dinger Equation is not valid for a photon,
but there is an equation that \textcolor{black}{has }the \textit{same}
form as the Schr�dinger Equation that \textit{is} valid. The resolution
to issue number 2 comes in two limits. In the limit of relativistic
massive particles on should use the relativistic Schr�dinger Equation,
in which the standard position observable is not valid \cite{Newton1949}.
That is, both photons and electrons do not have simple position observable
in the relativistic limit. In fact, there is an inherent contradiction
in the non-relativistic Schr�dinger Equation: an electron localized
to a point will contain infinite momentum components, and, hence,
will be relativistic. 

Consequently, the position observable in the Schr�dinger Equation
is only a valid approximation in certain limits. The same holds for
the photon; when one is in both the paraxial limit and in the limit
where the slowly varying frequency envelope amplitude approximation
is valid \cite{Deutsch1991,Aiello2005}, the photon wavefunction can
be written as $\Psi(\overset{\rightharpoonup}{p},\sigma)$, where
$\overset{\rightharpoonup}{p}=(p_{x},p_{y},p_{z})$ is the photon's
momentum and $\sigma=H,V$ is the polarization of the photon ($H$=horizontal,
$V$=vertical). In these limits, $p_{z}=hf/c$ and $z=t/c$, where
$f$ is the photon's frequency and $t$ is photon's time ($h$ is
Heisenberg's constant and $c$ is the speed of light). These parameters
are all defined relative to some time-space coordinate system. In
these limits, transverse positions $x$ and $y$; transverse momenta
$p_{x}$ and $p_{y}$; frequency $f$ (or $p_{z}$); and time $t$
(or $z$) can each be measured as an observable. Moreover, in these
limits, the photon wavefunction is normalized to unity in the same
manner as the electron wavefunction:
\begin{eqnarray*}
\sum_{\sigma}\int d\overset{\rightharpoonup}{p}\left|\Psi(\overset{\rightharpoonup}{p},\sigma)\right|^{2} & = & 1\\
\sum_{\sigma}\iiint dxdydz\left|\Psi(x,y,z,\sigma)\right|^{2} & = & 1\\
\sum_{\sigma}\iiint dp_{x}dp_{y}df\left|\Psi(p_{x},p_{y},f,\sigma)\right|^{2} & = & 1.
\end{eqnarray*}
These normalizations show some of the many equivalent ways of expressing
the photon wavefunction.

The relation to the second quantization of light is that the wavefunction
of the photon becomes a `mode' in which photons can be created such
that,
\[
\mathbf{a}_{\Psi}^{\dagger}=\sum_{\sigma}\iiint dp_{x}dp_{y}df\Psi(p,f,\sigma)\mathbf{a}_{p_{x},p_{y},f,\sigma}^{\dagger},
\]
where $\mathbf{a}_{p_{x},p_{y},f,\sigma}^{\dagger}$ is a photon creation
operator for the plane wave defined by $p_{x}$, $p_{y},$ $f$, and
$\sigma$. In this context, the single photon wavefunction is same
state as$\left|1\right\rangle _{\Psi}=\mathbf{a}_{\Psi}^{\dagger}\left|0\right\rangle $. 

In our experiment, we focus on the transverse position $x$ wavefunction
$\Psi(x)$ of the photon. This is valid if the full wavefunction can
be factorized, $\Psi(x,y,z,\sigma)=\Psi(x)\Psi(y)\Psi(z)\Psi(\sigma)$,
which we ensure by using photons of a single frequency, polarization
and spatial mode (the fiber mode). Since photons do not interact,
it is not crucial to send them through one at a time. This is true
as long as the only measurements are of the number operator $\mathbf{n}=\mathbf{a}^{\dagger}\mathbf{a},$
i.e. intensity measurements. Hence, for experimental simplicity we
use an attenuated laser as our source of photons.

\subsection{Back-Action in the measurement of the Dirac Distribution}

Here, we use back-action as a term to describe the disturbance induced
in the measured system by a measurement. Our theoretical discussion
of weak measurement described results in the limit of zero interaction
between the measurement apparatus (the photon polarization and waveplate
sliver) and the system (the photon's transverse position). While this
limit is unphysical, generally, the average result of a weak measurement
(the `weak value' or `weak average') converges to its limit-value
asymptotically as the interaction strength is reduced. Consequently,
the weak average is a good approximation even for reasonable interaction
strengths. In our measurement, we use both a two-dimensional operator
$\boldsymbol{\pi}_{x}$and pointer (i.e. polarization). This bounds
the effect of backaction and makes it simple to account for \cite{Kedem2010,Hofmann2012a}.
Consider the unitary that models our weak measurement of position,
\[
\mathbf{U}=e^{-i\varphi\boldsymbol{\sigma}_{y}\boldsymbol{\pi}_{x}},
\]
where $\boldsymbol{\sigma}_{y}$ is the Pauli operator and $\varphi$
is the strength of the interaction. Without any approximations, this
induces a rotation of the polarization by $\varphi$ if the photon
is at position $x$. As $\varphi$ increases, it will entangle the
polarization and position degrees of freedom. This entanglement is
the cause of a dephasing in the $x$-basis, i.e. the back-action.
In turn, this dephasing suppresses the Dirac distribution by an amount
proportional to $\mathrm{Prob}(x)=\left\langle \Psi\right|\boldsymbol{\pi}_{x}\left|\Psi\right\rangle $,
$\mathrm{\mathbb{\widetilde{P}}}(x,p)\rightarrow\mathrm{\mathbb{\widetilde{P}}}(x,p)-(1-\cos(\varphi))\mathrm{Prob}(x)$.
This effect creates a small offsets in every point of the Dirac distribution
that are difficult to see. Hence, we do not correct our Dirac distributions
for the back-action. On the other hand, if we examine the $x$-basis
density matrix found from the measured Dirac distribution, we find
that the back-action suppresses only the diagonal elements. We correct
our presented density matrices by dividing the diagonals by $\mathrm{cos}(\varphi$).
Note that since this correction grows as $1-\varphi^{2}/2$ for small
$\varphi$, one can measure a good approximation to the true weak
average even for experimentally reasonable values of$\varphi$. As
one reduces $\varphi$, at some point another source of experimental
error or uncertainty will become dominant. In practice, there is no
need to be in the limit $\varphi=0$.

\subsection{Properties of the Dirac-Distribution}

The Dirac distribution shares many of the properties of other the
other common quasi-probability distributions - the Q, P and Wigner
functions. Namely, any physically measurable property can be calculated
directly and simply from the Dirac distribution without first transforming
to the density matrix \cite{Agarwal1970a}.

1. It can be used to directly calculate the average result of measuring
an observable $\mathrm{A}$ on state $\boldsymbol{\rho}$: 
\begin{equation}
\left\langle A\right\rangle =\mathrm{Tr}\left[\mathrm{A}\boldsymbol{\rho}\right]=2\pi\iint dxdp\mathrm{\mathbb{\widetilde{P}}_{\rho}}(x,p)\mathrm{\mathbb{\widetilde{P}}_{A}^{*}}(x,p),
\end{equation}
where $\mathrm{\mathbb{\widetilde{P}}_{A}}(x,p)$ is the Dirac distribution
for observable $\mathbf{A}$ and {*} is the complex conjugate. 

2. With $\mathbf{A}$ chosen to be the identity operator $\mathbf{I}$,
we find that $\iint dxdp\mathrm{\mathbb{\widetilde{P}}_{\rho}}(x,p)=\mathrm{Tr}\left[\boldsymbol{\rho}\right]=1$.
In other words, the Dirac distribution is normalized in the same manner
as a probability distribution. 

Notice that Eq. 4 in the main paper shows that once the strength of
the measurement (i.e.$\sin\varphi$) is accounted for, the measured
Dirac distribution requires no further normalization (unlike the direct
wavefunction measurement in Ref. \citep{Lundeen2011}).

3. With $\mathbf{A}=\mathbf{\rho}$, we find the purity $\mu=2\pi\iint dxdp\left|\mathrm{\mathbb{\widetilde{P}}_{\rho}}(x,p)\right|^{2}$,
which reaffirms that purity is a global property of the density operator
and thus, we are unable to measure purity without completely \textcolor{black}{determining
}$\boldsymbol{\rho}$. 

4. And finally, with $\mathbf{A}=\left\vert x\right\rangle \left\langle x\right\vert $
or $\left\vert p\right\rangle \left\langle p\right\vert $, we see
that the $x$ and $p$ marginals are equal to the probability distributions
of outcomes $x$ and $p$, e.g. $\int dx\mathrm{\mathbb{\widetilde{P}}_{\rho}}(x,p)=\mathrm{Prob}(p)$. 

Consequently, the result of our joint weak measurement, the Dirac
distribution of the density operator, is a capable alternative to
the standard quantum quasi-probability distributions, such as the\textcolor{black}{{}
Wigner }function. Its peculiarity is that it is complex whereas\textcolor{black}{{}
classical }probabilities are real. Nonetheless, our method for directly
measuring it provides an operational meaning to both its real and
imaginary parts; they appear right on our measurement apparatus, in
the shifts in the two conjugate observables of the pointer, e.g.,
$x$ and $p$.

\subsection{Bayes' Law and the Dirac Distribution}

To compare with theory, we determine the Bayesian propagator by calculating
the overlaps of the constituent eigenstates, $\mathrm{\mathrm{\mathbb{\widetilde{P}}}}(k'|x,p)=\left\langle p|k'\right\rangle \left\langle k'|x\right\rangle /\left\langle p|x\right\rangle $.
We use the complex conditional probability,
\begin{widetext}
\[
\mathrm{\mathrm{\mathbb{\widetilde{P}}}}(k'|x,p)=R\cdot\frac{\exp\left[2\pi i\left(\sqrt{\Delta z^{2}+(x_{FT}-k')^{2}}/\lambda+\left(xk'-x_{FT}x\right)/(f_{FT}\lambda)+\alpha\right)\right]}{\sqrt{\Delta z^{2}+(x_{FT}-k')^{2}}}
\]
\end{widetext}
 where $\alpha=x\Delta z/(\lambda\sqrt{x^{2}+f_{FT}^{2}})$ and $R$
is a normalization \citep{Lohmann1993} (Here, we have left out the
effect of the magnification lens) With this, we propagate the experimental
Dirac distribution in Fig. 2 (b) to theoretically predict $\mathrm{\mathrm{\mathbb{\widetilde{P}}}}(x,k')=\sum_{p}\mathrm{\mathbb{\widetilde{P}}}(k'|x,p)\cdot\mathrm{\mathbb{\widetilde{P}}}(x,p)$
and compare to direct measurement of $\mathrm{\mathrm{\mathbb{\widetilde{P}}}}(x,k')$.

\appendix

%

\end{document}